\documentclass[amsmath,amssymb,12pt,preprint]{revtex4}

\usepackage{epsf}

\begin{document}

\title{Ln(3) and Black Hole Entropy\footnote{Contribution to the Proceedings of the 3rd International Symposium on Quantum Theory and Symmetries, Cincinnati, September 2003}}

\author{Olaf Dreyer}
\address{Perimeter Institute for Theoretical Physics\\
    35 King Street North\\
    Waterloo, Ontario\\
    Canada N2J 2W9\\
    E-mail: odreyer@perimeterinstitute.ca }
\date{January 14, 2004}

\begin{abstract}We review an idea that uses details of the quasinormal
mode spectrum of a black hole to obtain the Bekenstein-Hawking
entropy of $A/4$ in Loop Quantum Gravity. We further comment on a
recent proposal concerning the quasinormal mode spectrum of
rotating black holes. We conclude by remarking on a recent
proposal to include supersymmetry.
\end{abstract}

\maketitle

\section{Introduction}
Ever since Bekenstein\cite{b:ent} and Hawking\cite{Hawking:1975sw}
argued that a black hole has an entropy $S$ which is a quarter of
its horizon area $A$ it has been a challenge to theoretical
physicists to explain what this is an entropy of. In recent years
candidate theories of quantum gravity have provided concrete
realizations of quantum mechanical microstates. Both String
theory\cite{pol} and Loop Quantum Gravity\cite{thomas} have given
derivations for the Bekenstein-Hawking entropy.

In this article we will describe how one can fix a free parameter,
known as the Immirzi\cite{immirzi} parameter, to obtain the exact
Bekenstein-Hawking result for the entropy in Loop Quantum Gravity.
We will do this using an input from classical relativity, namely
the quasinormal mode spectrum of black holes. The first three
sections of this paper deal with this argument. The following
sections then deal with rotating black holes and also discuss a
new insight into the problem that relies on supersymmetry.

\section{Quasinormal Modes in the High Damping Limit}
\begin{figure}[ht]
\centerline{\epsfxsize=8cm\epsfbox{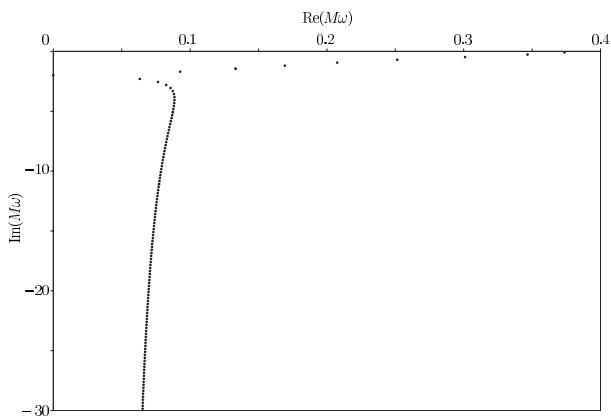}} \caption{The
position of the quasinormal mode frequencies in the complex
$\omega$-plane. For higher damping the real part of the frequency
becomes approaches the limiting value of $\ln 3/8 \pi M$.
\label{fig1}}
\end{figure}

The reaction of a black hole to perturbations is dominated by an
infinite discrete set of damped oscillation called quasinormal
modes\cite{nollertrev,kokkotas}. One remarkable feature of the
quasinormal mode spectrum for non-rotating black holes is that the
real part of the frequency approaches the non-zero
value\footnote{This fact was first observed by Hod\cite{hod} using
numerical data by Nollert\cite{nollert} and it has been recently
proved rigorously by Motl\cite{lubos1}.}
$\omega_\infty = \ln 3/8 \pi M$ as the damping increases (see
figure \ref{fig1}).

A non-rotating black hole has thus associated with it one
distinguished frequency. One might then ask if there is a special
process in the underlying quantum theory that gives rise to this
particular frequency. We will come back to this question after
having taken a closer look at Loop Quantum Gravity.

\section{Facts from Loop Quantum Gravity}
In Loop Quantum Gravity a surface is thought to acquire area
through the punctures of spin network edges. Spin networks form
the basis of the Hilbert space of LQG. They consist of graphs
whose edges are labelled by representations of the gauge group of
the theory. In our case this is SU(2) (or SO(3)). The labels are
thus (half-)integers. For a puncture with label $j$ the area is
\begin{equation}\label{eqn:area}
  A(j) = 8\pi l_P^2 \gamma \sqrt{j(j+1)}.
\end{equation}
Here $l_P^2$ is the Planck length. The free parameter $\gamma$ is
the so-called Immirzi\cite{immirzi} parameter.

One can think of the horizon area of a black hole to be the result
of a large number of spin network edges puncturing the horizon
surface. The Hilbert space of the theory living on the boundary is
increased by each puncture. If the edge has a label $j$ the
dimension of the Hilbert space increases by a factor of $2j+1$.
The statistically most important contribution comes from the
lowest possible spin $j_{\min}$. The dimension of the Hilbert
space is thus
\begin{equation}
  (2 j_{\min}+1)^N,
\end{equation}
where $N$ is the number of punctures. Since the area due to one
puncture is $A(j_{\min})$ (see formula (\ref{eqn:area})) this
number $N$ is given by $A/A(j_{\min})$. The entropy of the black
hole is thus
\begin{equation}
  S = \frac{A}{8\pi
  l_P^2\gamma\sqrt{j_{\min}(j_{\min}+1)}}\ln(2j_{\min}+1).
\end{equation}
The entropy is proportional to the area but the result so far is
not satisfactory since it contains the free parameter $\gamma$.

\section{Fixing the Immirzi Parameter}\label{sec:fix}
\begin{figure}[ht]
\centerline{\epsfxsize=8cm\epsfbox{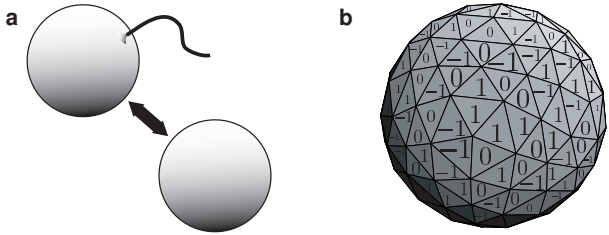}}
\caption{\textbf{\textsf{a}} The proposed quantum transition
giving rise to the emission of a quantum of energy
$\hbar\omega_\infty$. \textbf{\textsf{b}} The emerging ``It from
trit" picture.\label{fig2}}
\end{figure}

When discussing the high damping limit of the quasinormal mode
frequencies we asked the question whether there could exist a
quantum mechanical process that could give rise to the the
limiting frequency $\omega_\infty$. Given the picture developed in
the last section such a process is easy to imagine. It is the
appearance or disappearance of a puncture of the horizon (see
figure \ref{fig2}\textbf{\textsf{a}}).

As Hod\cite{hod} pointed out, the emission of a quanta of
frequency $\omega_\infty$ would change the area of a non-rotating
black hole by $4\ln 3\; l_P^2$. This has to be equated with the
change in area $A(j_{\min})$ due to one puncture. This equation
fixes the Immirzi parameter and gives for the entropy
\begin{equation}\label{eqn:sfinal}
  S = \frac{A}{4 l_P^2}\frac{\ln (2j_{\min} +1)}{\ln 3}.
\end{equation}
We see that we obtain exactly the Bekenstein-Hawking result if
$j_{\min}=1$. This can be achieved if one chooses SO(3) as the
gauge group instead of SU(2)\cite{dreyer}. Wheelers\cite{wheeler}
``It from bit'' gets thus modified to ``It from trit'' (see figure
\ref{fig2}b)\footnote{I thank Jonathan Oppenheim for this quote.}.

\section{Rotating Black Holes}
In the last section we have used quasinormal modes to learn
something about the quantum mechanics of non-rotating black holes.
A new argument by Hod\cite{hod2} tries to go the other way. This
time the quantum mechanics of black holes is used to gain insight
into the quasinormal mode spectrum of a rotating black hole. The
starting point is the first law of black hole mechanics
\begin{equation}\label{eqn:flaw}
  \Delta M = T \Delta S + \Omega \Delta J.
\end{equation}
Using the expressions in the Kerr metric for the temperature $T =
\sqrt{M^2 - a^2}/4\pi M r_+$ and the angular velocity $\Omega =
a/(r_+^2 + a^2)$, where $r_\pm = M \pm \sqrt{M^2-a^2}$ are the
horizon radii, together with the identification $\Delta M =
\hbar\;\text{Re}(\omega)$ and $\Delta S = 4 \ln(3)$ one obtains an
expression for Re$(\omega)$ from equation (\ref{eqn:flaw}):
\begin{equation}
    \text{Re}(\omega) = \frac{\ln(3)}{\pi} \frac{(M^2 -%
    a^2)^{1/2}}{M(M + (M^2 - a^2)^{1/2})}%
     + \frac{2 a}{(M + (M^2 -a^2)^{1/2})^2 + a^2}\label{eqn:omegaofa}
\end{equation}
We thus obtain a prediction for how the frequency Re$(\omega)$
changes as a function of the rotation parameter $a=J/M$.

\begin{figure}[ht]
\centerline{\epsfxsize=8cm\epsfbox{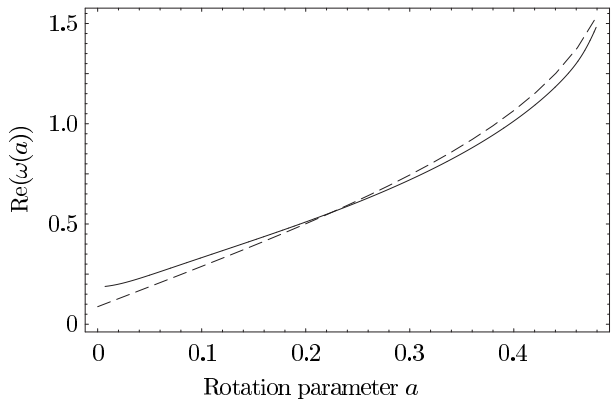}} \caption{Comparing
Hod's prediction with numerical data for the $n=8$ and $l=m=2$
modes. The dashed line is the theoretical prediction. The curve is
matched well. The biggest discrepancy occurs at the origin, which
is to be expected since we are looking at small damping and the
$\omega$ is assumes its limiting value for large damping.
\label{fig3}}
\end{figure}

To check the validity of this prediction Hod compared it to
numerical data obtained by Onozawa\cite{ono}\footnote{%
Since this talk was given our knowledge of quasinormal modes for the Kerr black hole has improved dramatically. The asymptotic behavior has been exhaustively investigated numerically by Berti et. al. \cite{berti1,berti2}. The results contained in these references show clearly that the prediction made by Hod is not valid.%
}, who used numerical
techniques developed by Leaver\cite{leaver}. The graph obtained is
reproduced in figure \ref{fig3} using our own numerical data.

The two graphs match well. The largest deviation occurs at the
origin. This is to be expected since we are looking at low
damping.

\begin{figure}[ht]
\centerline{\epsfxsize=12cm\epsfbox{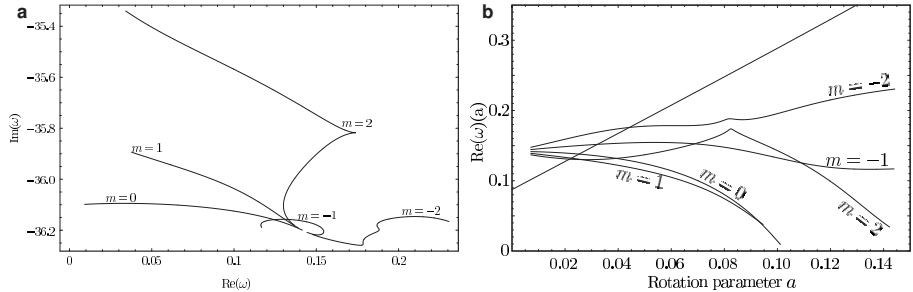}}
\caption{\textbf{\textsf{a}} The Kerr quasinormal
    modes for higher damping. We have $n=79$, $l=2$, $m=-2,\ldots,2$,
    and $a = 0 \ldots\approx 0.14$.\textbf{\textsf{b}} The function
    Re$(\omega)(a)$. There is no good agreement for any of the
    curves. \label{fig4}}
\end{figure}

It is assumed that the situation improves for larger damping. This
is not the case though for the largest damping we have tested.
Figure \ref{fig4}\textbf{\textsf{a}} shows quasinormal modes for
$n=79$. Figure \ref{fig4}\textbf{\textsf{b}} then shows the
Re$(\omega)(a)$ plot, this time including all the values of $m$.
None of the graphs shows a good matching with the theoretical
prediction.

\section{Supersymmetry}
In section \ref{sec:fix} we saw that we could get perfect
agreement of the loop quantum gravity calculation with the
Bekenstein-Hawking result provided we change the gauge group from
SU(2) to SO(3). This step is problematic since it is not clear how
to include fermions in such a theory. Using supersymmetric spin
networks Ling and Zhang\cite{ling} were able to circumvent this
problem. For supersymmetric spin networks the area spectrum is
$A(j) = 8\pi\gamma l_P^2\sqrt{j(j+1/2)}$ and the dimension of the
representation spaces is $4j+1$. Using the same arguments as above
Ling and Zhang obtain the following result for the entropy:
\begin{equation}
  S = \frac{A}{4}\frac{\ln(4j_{\min}+1)}{\ln(3)}
\end{equation}
Exact agreement is thus obtained for $j_{\min}=1/2$.

\section{Conclusions}
We have seen how the classical theory of quasinormal modes could
be used to fix an ambiguity in Loop Quantum Gravity. A more
sophisticated treatment of black holes in Loop Quantum Gravity
will be required to see how real this connection is.

The recent results by Hod concerning rotating black holes are
interesting but the numerical data is ambiguous. More numerical
data on highly damped Kerr modes is required to settle this
question.

The appearance of supersymmetry is surprising. The deeper
significance will only become clear when we better understand the
dynamic processes underlying the proposed connection.

\end{document}